\documentclass[twocolumn,showpacs,preprintnumbers,amsmath,amssymb]{revtex4}

\usepackage{graphicx}
\usepackage{dcolumn}
\usepackage{bm}

\begin{document}

\preprint{Submitted to PRL, Revised Dec. 9}

\title{First Results from KamLAND: Evidence for Reactor Anti-Neutrino Disappearance}

\author{
K.~Eguchi,$^1$
S.~Enomoto,$^1$
K.~Furuno,$^1$
J.~Goldman,$^1$
H.~Hanada,$^1$
H.~Ikeda,$^1$ 
K.~Ikeda,$^1$
K.~Inoue,$^1$
K.~Ishihara,$^1$
W.~Itoh,$^1$
T.~Iwamoto,$^1$
T.~Kawaguchi,$^1$
T.~Kawashima,$^1$
H.~Kinoshita,$^1$
Y.~Kishimoto,$^1$
M.~Koga,$^1$
Y.~Koseki,$^1$
T.~Maeda,$^1$
T.~Mitsui,$^1$
M.~Motoki,$^1$
K.~Nakajima,$^1$
M.~Nakajima,$^1$
T.~Nakajima,$^1$
H.~Ogawa,$^1$
K.~Owada,$^1$
T.~Sakabe,$^1$
I.~Shimizu,$^1$
J.~Shirai,$^1$
F.~Suekane,$^1$
A.~Suzuki,$^1$
K.~Tada,$^1$
O.~Tajima,$^1$
T.~Takayama,$^1$
K.~Tamae,$^1$
H.~Watanabe,$^1$
J.~Busenitz,$^2$
\v{Z}.~Djur\v{c}i\'{c},$^2$
K.~McKinny,$^2$
D-M.~Mei,$^2$
A.~Piepke,$^2$
E.~Yakushev,$^2$
B.E.~Berger,$^3$
Y.D.~Chan,$^3$
M.P.~Decowski,$^3$
D.A.~Dwyer,$^3$
S.J.~Freedman,$^3$
Y.~Fu,$^3$
B.K.~Fujikawa,$^3$
K.M.~Heeger,$^3$
K.T.~Lesko,$^3$
K.-B.~Luk,$^3$
H.~Murayama,$^3$
D.R.~Nygren,$^3$
C.E.~Okada,$^3$
A.W.P.~Poon,$^3$
H.M.~Steiner,$^3$
L.A.~Winslow,$^3$
G.A.~Horton-Smith,$^4$
R.D.~McKeown,$^4$
J.~Ritter,$^4$
B.~Tipton,$^4$
P.~Vogel,$^4$
C.E.~Lane,$^5$
T.~Miletic,$^5$
P.W.~Gorham,$^6$
G.~Guillian,$^6$
J.G.~Learned,$^6$
J.~Maricic,$^6$
S.~Matsuno,$^6$
S.~Pakvasa,$^6$
S.~Dazeley,$^7$
S.~Hatakeyama,$^7$
M.~Murakami,$^7$
R.C.~Svoboda,$^7$
B.D.~Dieterle,$^{8}$
M.~DiMauro,$^{8}$
J.~Detwiler,$^{9}$
G.~Gratta,$^{9}$
K.~Ishii,$^{9}$
N.~Tolich,$^{9}$
Y.~Uchida,$^{9}$
M.~Batygov,$^{10}$
W.~Bugg,$^{10}$
H.~Cohn,$^{10}$
Y.~Efremenko,$^{10}$
Y.~Kamyshkov,$^{10}$
A.~Kozlov,$^{10}$
Y.~Nakamura,$^{10}$
L.~De~Braeckeleer,$^{11}$
C.R.~Gould,$^{11}$
H.J.~Karwowski,$^{11}$
D.M.~Markoff,$^{11}$
J.A.~Messimore,$^{11}$
K.~Nakamura,$^{11}$
R.M.~Rohm,$^{11}$
W.~Tornow,$^{11}$
A.R.~Young,$^{11}$
and Y-F.~Wang$^{12}$\\
(KamLAND Collaboration)
}

\affiliation{
$^1$ Research Center for Neutrino Science, Tohoku University, Sendai
980-8578, Japan \\
$^2$ Department of Physics and Astronomy, University of Alabama, Tuscaloosa,
Alabama 35487, USA \\
$^3$ Physics Department, University of California at Berkeley and Lawrence Berkeley National
Laboratory, Berkeley, California 94720, USA \\
$^4$ W.~K.~Kellogg Radiation Laboratory, California Institute of
Technology, Pasadena, California 91125, USA \\
$^5$ Physics Department, Drexel University, Philadelphia, Pennsylvania
19104, USA \\
$^6$ Department of Physics and Astronomy,
University of Hawaii at Manoa, Honolulu, Hawaii 96822, USA \\
$^7$ Department of Physics and Astronomy, Louisiana State University, Baton Rouge, Louisiana
70803, USA \\
$^{8}$ Physics Department, University of New Mexico, Albuquerque,
New Mexico 87131, USA \\
$^{9}$ Physics Department, Stanford University, Stanford, California
94305, USA \\
$^{10}$ Department of Physics and Astronomy, University of Tennessee, Knoxville, Tennessee
37996, USA \\
$^{11}$ Triangle Universities Nuclear Laboratory, Durham, North Carolina 27708, USA
and \\
Physics Departments at Duke University, 
North Carolina State University, and 
 the University of North Carolina at Chapel Hill \\
$^{12}$ Institute of High Energy Physics, Beijing 100039, People's Republic of China 
}%

\date{December 9, 2002}

\begin{abstract}

KamLAND has been used to measure the flux of $\bar{\nu}_e$'s  from  distant nuclear
reactors. In an exposure of 162 ton$\cdot$yr (145.1 days) the ratio of the
number of observed inverse $\beta$-decay events to the expected number of
events without disappearance is $0.611\pm 0.085 {\rm (stat)} \pm 0.041 {\rm (syst)} $
for $\bar{\nu}_e$ energies $>$ 3.4 MeV. The deficit of events is inconsistent with the expected rate for standard $\bar{\nu}_e$
propagation at the 99.95\% confidence level. In the context of two-flavor
neutrino oscillations with CPT invariance, these results exclude all 
oscillation solutions but the `Large Mixing Angle'
solution to the solar neutrino problem using reactor $\bar{\nu}_e$ sources.

\end{abstract}

\pacs{14.60.Pq, 26.65.+t, 28.50.Hw, 91.65.Dt}

\maketitle

%
 
The primary goal of the Kamioka Liquid scintillator Anti-Neutrino Detector
(KamLAND) experiment~\cite{kamland} is a search for the oscillation of $\bar{\nu}_e$'s emitted from distant power reactors. The long baseline, typically 180~km, enables KamLAND to address the oscillation solution of the `solar neutrino 
problem' using reactor anti-neutrinos under laboratory conditions.
The  inverse $\beta$-decay reaction, $\bar{\nu}_e +
p\rightarrow e^+ + n$, is utilized to detect
$\bar{\nu}_e$'s  with energies above 1.8 MeV in liquid scintillator (LS) \cite{RMP}. 
The detection of the $e^+$ and the 2.2~MeV $\gamma$-ray from neutron capture
on a proton in delayed coincidence
is a powerful tool for reducing background.
This letter presents the first results from an analysis of 162 ton$\cdot$yr of
the reactor $\bar{\nu}_e$ data. 

\begin{figure}
\includegraphics[angle=270,width=9cm]{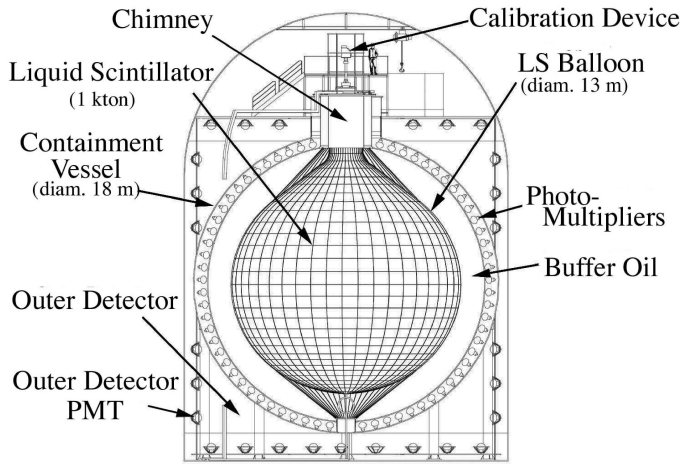}
\caption[]{ Schematic diagram of the KamLAND detector.}
\label{fig:detector}
\end{figure}

%

KamLAND is located at the
site of the earlier Kamiokande~\cite{Kamioka}, with an average
rock overburden of 2,700 m.w.e. resulting in 0.34 Hz of
cosmic-ray  muons in the detector volume. 
As shown in Fig.~\ref{fig:detector},
the neutrino detector/target is 1 kton of ultra-pure LS
contained in a 13-m-diameter spherical balloon made of
135-$\mu m$-thick transparent nylon/EVOH
(Ethylene vinyl alcohol copolymer)
composite film.
The balloon is supported and constrained by a network of kevlar ropes.
The LS is 80\%  dodecane, 20\%  pseudocumene (1,2,4-Trimethylbenzene), 
and 1.52 g/liter of PPO (2,5-Diphenyloxazole)
as a fluor. 
A buffer of dodecane and isoparaffin oils between 
the balloon and an 18-m-diameter spherical stainless-steel
containment vessel shields the LS from external radiation.
During the filling procedure
a water extraction and nitrogen bubbling method \cite{purify}, optimized for KamLAND,
was used to purify the LS and buffer oil; PPO prepurification and dust removal were especially important. 

The specific gravity of the buffer oil is adjusted to be 0.04\% lower
than that of the LS.
An array of 1,879
photomultiplier tubes (PMTs), mounted on the inner surface of the 
containment vessel, completes the
inner detector (ID) system. This array includes 1,325 specially developed 
fast PMTs with
17-inch-diameter photocathodes and 554 older Kamiokande 20-inch 
PMTs \cite{Kamiokande}. 
While the total photo-cathode coverage is 34\%,
only the 17-inch PMTs corresponding to 22\%  coverage
are used for the analysis in this letter. A 3-mm-thick acrylic
barrier at 16.6-m diameter helps prevent radon emanating from 
PMT glass from entering the LS. 
The containment vessel is  surrounded by a 3.2~kton 
water-Cherenkov detector with 225 20-inch PMTs. 
This outer detector (OD) 
absorbs
$\gamma$-rays and
neutrons from the surrounding rock and provides a tag for cosmic-ray 
muons. 
The primary ID trigger threshold is set at 200 PMT hits,
corresponding to about 0.7~MeV.
This threshold is lowered to 
120 
hits for 1 msec after the
primary trigger to observe lower energy delayed
activity. The OD trigger threshold is set to provide
$> 99\%$ tagging efficiency.

%

\begin{figure}
\includegraphics[angle=0,width=8cm]{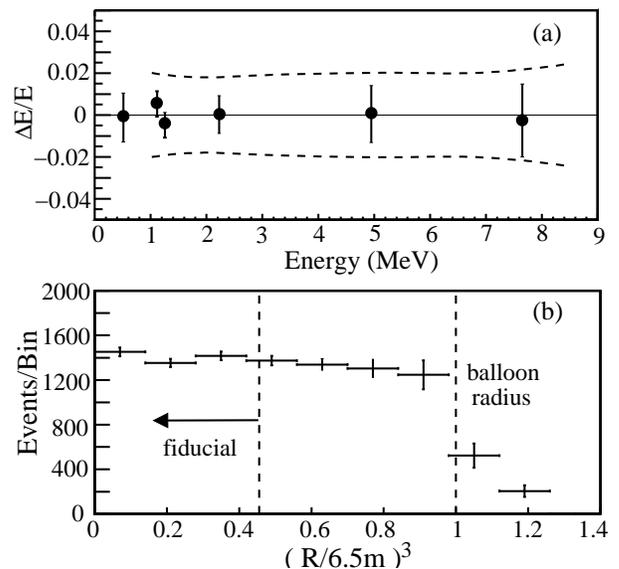}
%
\caption[]{(a) The fractional difference of the reconstructed average $\gamma$ energies and
average source energies. The dashed line shows the systematic error. (b) The $R^3$ vertex distribution of 2.2 MeV neutron capture $\gamma$'s.
}
\label{fig:calibration}
\end{figure}

%
The inner detector is calibrated with
$\gamma$-ray sources of ${}^{68}$Ge, ${}^{65}$Zn, ${}^{60}$Co, and Am-Be,
deployed at various positions along the vertical axis. 
These sources provide calibration energies in the 0.5 to 7.6 MeV region.
Energy measurements are based on the number of 
detected photoelectrons (p.e.) with corrections for PMT gain variation, solid angle, 
the density of PMTs, shadowing by suspension ropes, and light transparencies 
of the LS and buffer oil.
At the center, about 300 p.e. per MeV are observed.
Fig.~\ref{fig:calibration}~(a) shows the fractional deviation of the reconstructed
energies from the source energies. The
${}^{68}$Ge and ${}^{60}$Co sources emit two coincident $\gamma$-rays and are plotted at
an average energy
in Fig.~\ref{fig:calibration}~(a).
The observed energy resolution is $\sim$7.5\%/$\sqrt{E({\rm
MeV})}$.

The energy calibration is augmented with 
studies of ${}^{40}$K
and ${}^{208}$Tl (which are contaminants in the detector),
Bi-Po sequential decays, the spallation products $^{12}$B and 
${}^{12}$N, and $\gamma$'s
from thermal neutron captures on protons and $^{12}$C. 
The reconstructed
energy varies by less than 0.5\% within the
10-m-diameter 
fiducial volume; local variations near the chimney region are about 
1.6\%. 
The energy scale exhibits less than 0.6\% variation in time during the
entire data run. 
Corrections for quenching and Cherenkov light production are included, and contribute
to the systematic error in Fig.~\ref{fig:calibration} (a).
The estimated systematic error for the energy scale is 1.9\% at the 2.6~MeV
energy threshold which
corresponds to a 2.1\% uncertainty in the detected neutrino rate.
Correlated decays from Bi-Po and ${}^8$He/${}^9$Li 
yield energies between 0.6 MeV and $\sim$15 MeV and are used to extend the range of the
energy calibration.

The source positions are reconstructed from the relative times of 
PMT hits.
Energy-dependent radial adjustments are used
to reproduce
the known source positions to $\sim$5~cm; the
typical position reconstruction resolution is 25~cm.
Vertex reconstruction performance 
throughout the LS volume is verified by 
reproducing the uniform distribution of
2.2~MeV capture $\gamma$'s from spallation neutrons,
as shown in Fig.~\ref{fig:calibration} (b).

The data presented in this letter were obtained
during a period from March 4 through October 6, 2002. 
In total we collected 370 million events
in 145.1 days of live time, corresponding to an average
trigger rate of $\simeq$30 Hz. 
Events with fewer than
10,000 p.e. ($\sim$30 MeV) and no prompt OD veto signal are classified as 
`reactor $\bar{\nu}_e$ candidates'; more energetic events
are treated as `muon candidates.'

%
The criteria for selection of $\bar \nu_e$ events are 
(1) a fiducial volume cut ($R < 5$ m), 
(2) a time correlation cut (0.5~$\mu$sec $< \Delta T <$ 660 $\mu$sec), 
(3) a vertex correlation cut ($ \Delta R <1.6$ m),
(4) a delayed energy window cut (1.8 MeV $< E_{delay} <$ 2.6 MeV), and 
(5) a cut on the delayed vertex position requiring it to be more than 1.2~m from
the central vertical axis (to eliminate background from 
thermometers deployed in this region to monitor LS temperature). 
The overall efficiency associated with criteria (2)-(5) combined with the effect of (1) on the delayedvertex is $(78.3 \pm 1.6)\%$. 

Positrons annihilate in the LS and the detected energy 
includes the kinetic energy plus twice the electron rest mass; thus
$e^+$ from $\bar{\nu}_e$ events produce
$E_{prompt}=E_{\bar{\nu}_e}- \bar E_n -0.8$~MeV,
where $\bar E_n$ is the average neutron recoil energy.
Anti-neutrinos emitted by $^{238}$U and $^{232}$Th decays in the
Earth, `geo-neutrinos' ($\bar{\nu}_{geo}$), contribute low-energy events
with $E_{prompt}< 2.49$~MeV. 
For example, model Ia of $\bar{\nu}_{geo}$ in \cite{geonue} predicts
about 9 $\bar{\nu}_{geo}$ events in our data set.
However the abundances of U and Th and their
distributions in the Earth are not well known.  
To avoid ambiguities from $\bar{\nu}_{geo}$'s 
we employ (6) a prompt energy cut,  
$E_{prompt}> 2.6$~MeV in the present analysis.

%

Low-energy $\gamma$-rays from $^{208}$Tl ($\simeq$3~Hz) 
entering from outside the balloon
are potential sources of background in the range up to 3~MeV
and are strongly suppressed by the fiducial volume cut (1). 
The fiducial volume is estimated using the uniform distribution of spallation neutron 
capture events
shown in Fig.~~\ref{fig:calibration} (b).
The ratio of these events in the fiducial volume to those in the total volume
agrees with the geometric fiducial fraction to within  4.06\%.
The same method is used for higher energy events, $^{12}$N, $^{12}$B $\beta$'s
following muon spallation, and agrees within 3.5\%.
Accounting for uncertainty in the LS total mass of 2.1\%, 
we estimate the total systematic uncertainty of the fiducial volume to be
4.6\%.
The density of the LS is 0.780~g/cm$^3$ at $11.5^\circ$C;
the hydrogen-to-carbon ratio is computed from the LS components to be 1.969 and
was verified by
elemental analysis ($\pm 2\%$).
The specific gravity of the LS is measured to 0.01\% precision using a
commercial density meter, and we assign an
additional 0.1\% systematic error due to the uncertainty in the LS temperature.
The 408 ton fiducial mass thus contains $3.46 \times 10^{31}$
free target protons.

The trigger efficiency was determined to be 99.98\% using LED light sources.
The combined efficiency of the
electronics, data acquisition, and event reconstruction was carefully studied
using time distributions of uncorrelated events from calibration $\gamma$ sources.
We find that this combined efficiency is greater than $99.98\%$.
The vertex fitter yields
$>99.9\%$ efficiency within 2~m of known source positions.
Using calibrated $^{60}$Co and $^{65}$Zn sources,
the overall efficiency was verified to within
the 3\% uncertainty in source strengths. The detection efficiency
of delayed events from the Am-Be source (4.4~MeV prompt $\gamma$ and 2.2~MeV
delayed neutron capture $\gamma$ within 1.6~m) was verified with
1\% uncertainty.

From studies of Bi-Po sequential decays, the equilibrium
concentrations of $^{238}$U and $^{232}$Th in the LS are
estimated to be
$(3.5 \pm 0.5) \times 10^{-18}$ g/g and $(5.2~\pm~0.8)~\times~10^{-17}$~g/g, 
respectively. The observed background energy spectrum constrains the 
$^{40}$K contamination to be less than $2.7\times 10^{-16}$ g/g.
The accidental background, obtained from the observed 
flat distribution in the delayed time window 0.020-20~sec, is
$0.0086 \pm 0.0005$ events for the present data set.

\begin{table}[htb]
\caption{Background summary. }
\label{tab:table1}
\begin{tabular}{lc}
\hline \hline
Background              & Number of events \\
\hline
Accidental                & $0.0086 \pm 0.0005$ \\
$^9$Li/$^8$He                 & $0.94 \pm 0.85$    \\
Fast neutron              & $<$ 0.5            \\
\hline
Total B.G. events          & $0.95\pm 0.99$ \\
\hline \hline
 &  \\
\end{tabular}
\end{table}
%
%
%
\begin{table}[b]
\caption{Estimated systematic uncertainties (\%). }
\label{tab:table2}
\begin{tabular}{llll}

\hline \hline
Total LS mass 		 	& 2.1 &\ \ \ \ Reactor power                  & 2.0 \\
Fiducial mass ratio             & 4.1 &\ \ \ \ Fuel composition             & 1.0 \\
Energy threshold                & 2.1 &\ \ \ \ Time lag                           	& 0.28   \\
Efficiency of cuts                 & 2.1 &\ \ \ \ $\nu$ spectra \cite{spectra}        & 2.5 \\
Live time                          	& 0.07 &\ \ \ \ Cross section \cite{crosssection}   & 0.2 \\
\hline
Total systematic error        &	 &	 & 6.4\% \\
\hline \hline

\end{tabular}
\end{table}

At higher energies, the background is dominated by spallation products from
energetic muons. We observe
$\sim$3,000 neutron events/day/kton.  We also expect 
$\sim$1,300 
events/day/kton \cite{hagner} for various unstable products. 

Single
neutrons are efficiently suppressed by a 2-msec veto following a muon, but 
care is required to avoid neutrons which mimic the $\bar{\nu}_e$ 
delayed coincidence signal.
Most fast neutrons are produced by energetic muons which pass through both the OD
and the surrounding rock.  
This background is evaluated by
detecting delayed coincidence events with a
prompt signal associated with
a muon detected only by the OD. As expected,
a clear concentration of events near the balloon edge is observed. 
The expected background inside the fiducial volume is
estimated by extrapolating the vertex position distribution and
considering an OD reconstruction efficiency of 92\%. 
To estimate the number of background events due to neutrons 
from the surrounding rock, the OD-associated rate
is scaled by the relative 
neutron production and neutron shielding properties of the relevant materials.
We estimate that the total fast neutron background is less than 0.5 events for the
        entire data set.

Most radioactive spallation products simply beta decay, and are effectively
suppressed by requiring a delayed neutron signal. 
Delayed neutron emitters
like $^8$He ($T_{1/2}=119$~msec) and $^9$Li (178~msec) 
are eliminated by two time/geometry cuts:
(a) a 2-sec veto is applied for the entire fiducial volume following a
showering muon (those with more than $10^6$ p.e., $\sim$3~GeV,  extra energy
deposition), (b) for the remaining muons, delayed events within 2~sec
and 3~m from a muon 
track are rejected.
The efficiency of these cuts is calculated from
the observed correlation of spallation neutrons with muon tracks. 
The remaining  $^8$He and $^9$Li background is
estimated to be $0.94 \pm 0.85$.
The dead time due to the spallation cuts is 11.4\%.
This method is checked by exploiting the time
distribution of the events after a detected muon to
separate the short-lived spallation-produced activities from   
$\bar \nu_e$ candidates. 
The uncorrelated
$\bar \nu_e$ event distribution has a characteristic time constant of  
$1/R_\mu \simeq 3$~sec, where $R_\mu$ is the incident muon rate. 
Spallation products have a much shorter time constant ($\sim$0.2~sec).
These methods agree to 3\% accuracy. As shown in Table~\ref{tab:table1}
the total number of expected background events is $0.95 \pm 0.99 $,
 where the fast neutron contribution is included in the error estimate.  
\begin{figure}[t]
\includegraphics[angle=0,width=8cm]{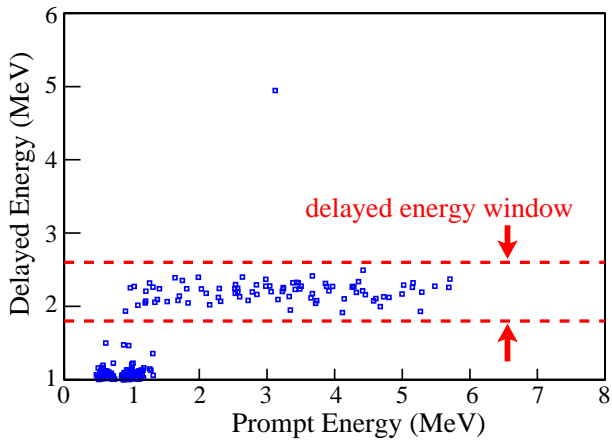}
\caption[]{ Distribution of $\bar{\nu}_e$ candidates with
 fiducial volume cut, time,  vertex correlation, and spallation cuts applied.
The prompt energy corresponds to the positron and the delayed energy to
the captured neutron.  The events within the horizontal lines bracketing the delayed
energy of 2.2~MeV are due to thermal neutron capture on protons.
The events with prompt energy below $\sim$0.7 MeV are obtained from the
 delayed trigger. The one event with delayed energy near 4.95~MeV is consistent with theexpected 0.54\% fraction from ${}^{12}$C$(n,\gamma)$.}

\label{fig:data}
\end{figure}

%

Instantaneous thermal power generation, burn-up and fuel
exchange records for all 
Japanese commercial power reactors are provided
by the power companies.
The time dependence of the thermal power generation data is checked
by comparison with the independent records of electric power generation. 
The fission rate for each fissile element is calculated from these
data, resulting in a systematic uncertainty in the $\bar{\nu}_e$ flux of
less than 1\%. Averaged over the present live-time period,
the relative fission yields from
the various fuel components are $^{235}$U : $^{238}$U :
$^{239}$Pu : $^{241}$Pu = 0.568 : 0.078 : 0.297 : 0.057. 
The $\bar{\nu}_e$ spectrum per fission and its error (2.48\%) are taken from the
literature \cite{spectra}. 
These neutrino spectra have been tested to a few percent
accuracy in previous
short-baseline reactor $\bar{\nu}_e$ experiments ~\cite{RMP,BUGEY}.
The finite $\beta$-decay lifetimes of fission products introduce an
additional uncertainty of 0.28\% to the $\bar{\nu}_e$ flux; 
this is estimated from the difference of the total $\bar{\nu}_e$ yield
associated with shifting the run time by one day.
The
contribution to the $\bar{\nu}_e$ flux from Korean reactors is
estimated to be (2.46 $\pm$ 0.25)\% from the reported electric power
generation rates.
Other reactors around the world give an average (0.70~$\pm$~0.35)\% contribution,
which is estimated by using reactor specifications from the 
International Nuclear Safety Center \cite{INSC}. 
The uncertainties for the event rate calculation  are summarized in Table 
~\ref{tab:table2}. 
The errors from reactors outside Japan are included in the table under
`Reactor Power'. 

Although the anti-neutrino flux at the location of KamLAND is due to many
nuclear reactors at 
a variety of distances, the $\bar{\nu}_e$ flux is actually dominated by a
few reactors at an average distance of $\sim$180~km.
More than 79\% of the computed flux arises from 26 reactors within
the distance range $138$-$214$~km. 
One reactor at 88~km contributes an additional 6.7\% to the flux 
and the other reactors are more than 295~km away.
This relatively narrow 
band of distances implies that for some oscillation parameters KamLAND can observe a
distortion of the $\bar{\nu}_e$
energy spectrum. 
 
The flux of anti-neutrinos from a reactor a distance $L$ from KamLAND  is approximately
proportional to the thermal power flux $ P_{th}/4\pi L^2$, where $P_{th}$ is the 
reactor thermal power.
The integrated total thermal power flux during the measurement live time is $
254 ~{\rm Joule/cm}^2$. The systematic error assigned to the thermal power is
conservatively
taken as 2\% from the regulatory specification for safe reactor operation.
The corresponding expected number of reactor neutrino events 
(in the absence of neutrino oscillations)
in the fiducial volume for this data set
is $86.8\pm 5.6$. 
%
\begin{figure}
\includegraphics[width=9cm]{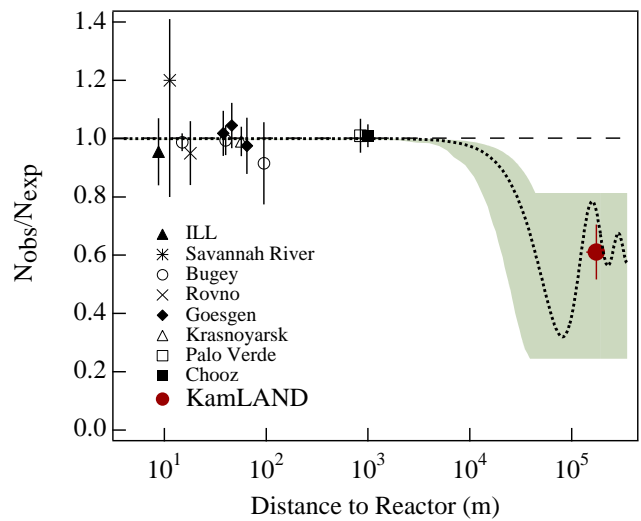}
\caption[]{The ratio of measured to expected $\bar \nu_e$ flux from reactor 
experiments~\cite{pdg}.  The solid dot is the KamLAND point plotted at a 
flux-weighted average distance (the dot size is indicative of the
spread in reactor distances).  The shaded region indicates the range of flux 
predictions corresponding to the 95\% C.L. LMA region found in a 
global analysis of the solar neutrino data \cite{solar}. The dotted 
curve corresponds to $\sin^2 2 \theta =0.833$ and $\Delta m^2= 5.5\times 
10^{-5}$~eV$^2$ \cite{solar} and is representative of recent best-fit 
LMA predictions while the dashed curve shows the case of small mixing 
angles (or no oscillation).
}
\label{fig:reactors}
\end{figure}
%

%

The distribution of prompt and delayed energies for the 
present sample before the energy cuts 
are applied is shown in Fig.~\ref{fig:data}. 
A clear cluster of events from the
2.2~MeV capture $\gamma$'s 
is observed.
One event with delayed energy around 5 MeV 
is consistent with a thermal
neutron capture $\gamma$ on $^{12}$C.
The space-time correlation of the
prompt and delayed events is in good agreement with expectations, and the 
observed mean neutron capture time is $188 \pm 23 ~ \mu$sec. 
After applying the prompt and delayed energy cuts, 54 events remain as
the final sample.   
The ratio of the number of observed reactor $\bar{\nu}_e$ events to
that expected in the absence of neutrino oscillations is
$$ \frac{N_{obs}-N_{BG}}{N_{expected}}=0.611\pm 0.085 {\rm (stat)} \pm
0.041 {\rm (syst)}.$$ 
The probability that the
KamLAND result is consistent with the 
no disappearance hypothesis is less than 0.05\%.
Fig.~\ref{fig:reactors} shows the ratio of measured to expected flux
for KamLAND as well as previous reactor experiments 
as a function of the average distance from the source. 

\begin{figure}
\includegraphics[angle=0,width=9cm]{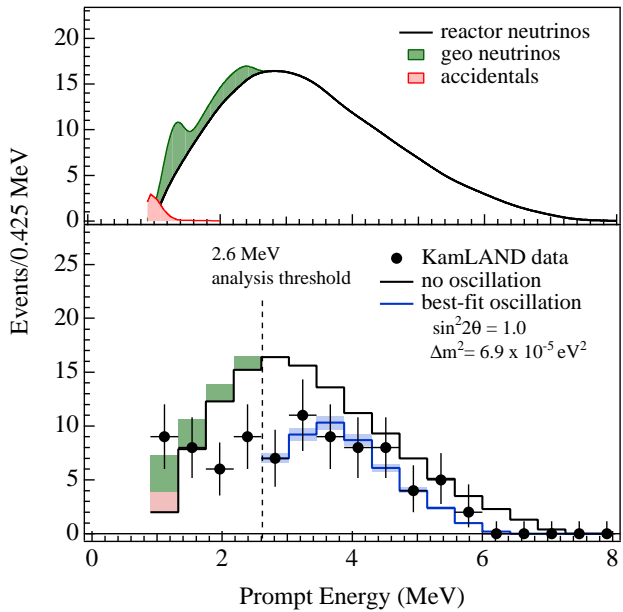}
\caption[]{ Upper panel: Expected reactor $\bar{\nu}_e$ energy spectrum with contributions
of $\bar{\nu}_{geo}$ (model Ia of \cite{geonue}) and accidental backround.
Lower panel: Energy spectrum of the observed prompt
 events (solid circles with error bars),
  along with the expected no oscillation spectrum (upper histogram, with $\bar{\nu}_{geo}$
and accidentals shown) and best
  fit (lower histogram) including neutrino
 oscillations. The shaded band indicates the 
systematic error in the best-fit spectrum.
The vertical dashed line corresponds to the 
analysis threshold at 2.6~MeV.
}
\label{fig:prompt}
\end{figure}

The observed prompt energy spectrum is shown
in Fig.~\ref{fig:prompt}. 
The expected positron
spectrum with no oscillations and the best
fit with two-flavor neutrino oscillations above the 2.6~MeV threshold are shown.
A clear deficit of events is observed.
The measured spectrum is consistent (93\% confidence) with a distorted spectrum shape
as expected from neutrino oscillations, but a renormalized 
no-oscillation shape is also consistent at 
53\% confidence. 

\begin{figure} 
\includegraphics[width=8cm]{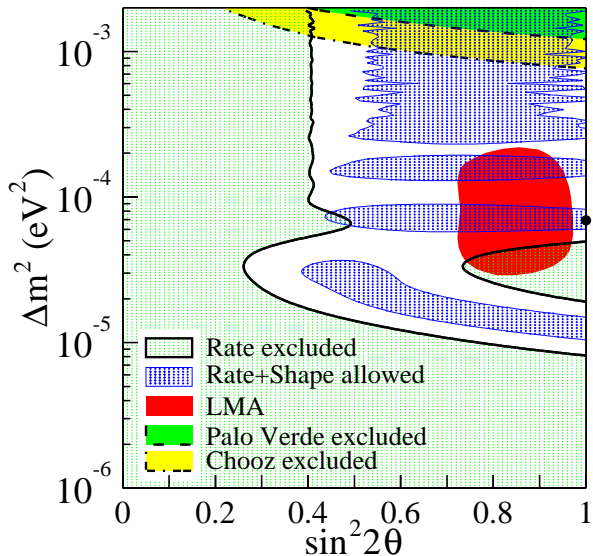}
\caption[]{ 
Excluded regions of neutrino oscillation parameters for the rateanalysis and allowed regions for the combined rate and shape analysis from KamLAND at 95\% C.L.
At the top are the 95\% C.L. excluded region from CHOOZ \cite{CHOOZ}
and Palo Verde \cite{Palo} experiments, respectively.  The 95\% C.L. allowed
region of the `Large Mixing Angle' (LMA) solution of solar neutrino
experiments \cite{solar} is also shown. The thick dot indicates the best fit to
the KamLAND data in the physical region: $\sin^2 2\theta = 1.0$ and
$\Delta m^2 = 6.9 \times 10^{-5}$eV$^2$.  All regions look identical
under $\theta \leftrightarrow (\pi/2-\theta)$ except for the LMA region.
}
\label{fig:contour}
\end{figure}

%
The neutrino oscillation parameter region for two-neutrino mixing is shown in
Fig.~\ref{fig:contour}.
The dark shaded area is the LMA region at 95\% C.L. derived from
~\cite{solar}.
The shaded region outside the solid line is excluded at 95\%
C.L. from the rate analysis with $\Delta \chi^2 \ge 3.84$ and
\begin{equation}
\chi ^2 = \frac{(0.611 - R(\sin^22\theta,\Delta m^2))^2}{0.085^2+0.041^2} \nonumber .
\end{equation}
Here, $R(\sin^22\theta,\Delta m^2)$ is the expected ratio with the 
oscillation parameters. 

The spectrum of the final event sample is then analyzed with a
maximum likelihood method to obtain the optimum set of oscillation parameters
with the following $\chi^2$ definition:
\begin{eqnarray}
\chi^2 & =  & \chi^2_{\rm rate}(\sin^22\theta,\Delta m^2,N_{\rm BG1\sim 2},\alpha_{\rm 1 \sim 4}) \nonumber \\
 & & -2\log L_{\rm shape}(\sin^22\theta,\Delta m^2,N_{\rm BG1\sim 2},\alpha_{\rm 1\sim 4}) \nonumber \\
 & & +\chi^2_{\rm BG}(N_{\rm BG1\sim 2})+\chi^2_{\rm distortion}(\alpha_{\rm 1\sim 4})\nonumber ,
\end{eqnarray}
where $L_{\rm shape}$ is the likelihood function of the spectrum including
deformations from various parameters.
$N_{\rm BG1\sim 2}$ are the estimated number of $^9$Li and $^8$He backgrounds
and $\alpha_{\rm 1\sim 4}$ are the parameters for the shape deformation 
coming from energy scale, resolution, $\bar{\nu}_e$ spectrum and fiducial volume.
These parameters are varied to minimize the $\chi^2$ at each pair of 
$\left[\Delta m^2, \sin^2\theta\right]$
with a bound from $\chi^2_{\rm BG}(N_{\rm BG1\sim 2})$ and 
$\chi^2_{\rm distortion}(\alpha_{\rm 1\sim 4})$.
The best fit
to the KamLAND data in the physical region
yields $\sin ^22\theta=1.0 $ and $\Delta m^2=6.9 \times 10^{-5}~\rm{eV}^2$
while the global minimum occurs slightly outside of the physical region
at $\sin ^22\theta=1.01$ with the same $\Delta m^2$.
These numbers can be compared to the best fit LMA values of 
 $\sin ^22\theta=0.833$ and $\Delta m^2=5.5\times 10^{-5}~\rm{eV}^2$ from
~\cite{solar}. 
The 95\% C.L. allowed regions from the 
spectrum shape analysis
are shown in Fig.~\ref{fig:contour}.
The allowed regions displayed for KamLAND correspond to $0 < \theta < {\pi \over 4}$
consistent with the solar LMA solution, but
for KamLAND the 
allowed regions in ${\pi \over 4} < \theta < {\pi \over 2}$ are identical \cite{hitoshi}.

Another spectral shape analysis is performed with a lower prompt
energy threshold of 0.9 MeV in order to check the stability of the above result 
and study the sensitivity to $\bar{\nu}_{geo}$.   
With this threshold, the total background is estimated to be $2.91
\pm 1.12$ 
events, most of which come from accidental and spallation
events.
The systematic error is 6.0\%, which is smaller than that for the final 
event sample due to the absence of an energy threshold effect. 
When the maximum likelihood is calculated, the
$\bar{\nu}_{geo}$ fluxes from $^{238}$U and $^{232}$Th are treated as free 
parameters.
The best fit in this analysis yields $\sin ^22\theta=0.91$ and $\Delta m^2=6.9\times
10^{-5}\rm{eV}^2$. 
These results and the allowed region of the oscillation parameters are
in good agreement with the results obtained above.   The
numbers of $\bar{\nu}_{geo}$ events for the best fit are 4 for $^{238}$U and 
5 for $^{232}$Th, which corresponds to $\sim$40 TW radiogenic heat generation
according to model Ia in~\cite{geonue}.
However, 
for the same model, $\bar{\nu}_{geo}$ production powers
from 0 to 110 TW are still allowed at 95\% C.L. with the same oscillation
parameters. 

If three neutrino generations are considered, the
$\bar{\nu}_e$ survival probability depends on two mixing
angles $\theta_{12}$ and $\theta_{13}$. In the region close to the best fit
KamLAND solution the survival probability
is, to a very good approximation,  given by
\begin{equation}
P(\bar{\nu}_e \rightarrow \bar{\nu}_e) \cong
\cos^4\theta_{13} \left[1 - \sin^2 2\theta_{12}\sin^2\frac{\Delta m_{12}^2
L}{4E_{\nu}} \right] \nonumber \> ,
\end{equation}
with $\Delta m_{12} \cong \Delta m^2$ from the 2-flavor analysis above.
The CHOOZ experiment \cite{CHOOZ} established an upper limit of
$\sin^2 2\theta_{13} < 0.15$, or $\cos^4 \theta_{13} \ge
0.92$. 
The best fit KamLAND result would correspond approximately to $0.86 <
\sin^2 2\theta_{12} < 1.0$.

%

 In summary KamLAND has used measurements
at large distances ($\sim$180~km) to demonstrate, for the first time,
reactor $\bar{\nu}_e$ disappearance at a high confidence level ($99.95\%$).
Since one expects a negligible reduction of $\bar \nu_e$ flux from the
SMA, LOW and VAC solar neutrino solutions, the LMA region is the only remaining 
oscillation solution
consistent with the KamLAND result and CPT invariance.
The allowed LMA region is further reduced by these 
results. Future measurements with greater statistical precision and reduced systematic errors
will enable KamLAND to provide a high-precision measurement of the neutrino 
oscillation parameters.

%

The KamLAND experiment is supported by the COE program of 
Japanese Ministry of Education, Culture, Sports, Science and Technology,
and funding from the 
United States 
Department of Energy. 
The reactor data are provided by courtesy of the following electric
associations in Japan; 
Hokkaido, 
Tohoku, 
Tokyo, 
Hokuriku, 
Chubu, 
Kansai, 
Chugoku, 
Shikoku and
Kyushu Electric Power Companies,  
Japan Atomic Power Co. and
Japan Nuclear Cycle Development Institute. 
Kamioka Mining and Smelting Company has provided service for activities
in the mine. 

%
%


\end{document}